\documentclass[prl,aps,twocolumn,noshowpacs,noshowkeys,superscriptaddress]{revtex4-1}
\usepackage{amsfonts}
\usepackage{amsmath}
\usepackage{amssymb}
\usepackage{graphicx}

\usepackage{textcomp}%
\newcommand{\vv}[1]{\mathbf{#1}}

\usepackage{color}

\begin{document}
\def\neel{Institut N\'{e}el, Universit\'{e} Grenoble Alpes - CNRS:UPR2940, 38042 Grenoble, France}
\def\ilm{Institut Lumi\`{e}re Mati\`{e}re, UMR5306, CNRS - Universit\'{e} Claude Bernard Lyon 1, 69622 Villeurbanne, France}
\author{Laure Mercier de L\'{e}pinay}
\affiliation{\neel}
\author{Benjamin Pigeau}
\affiliation{\neel}
\author{Benjamin Besga}
\affiliation{\neel}
\author{Pascal Vincent}
\affiliation{\ilm}
\author{Philippe Poncharal}
\affiliation{\ilm}
\author{Olivier Arcizet}
\affiliation{\neel}
\email{olivier.arcizet@neel.cnrs.fr}

\title{Universal Vectorial and Ultrasensitive Nanomechanical Force Field Sensor}

\maketitle

{\bf Miniaturization of force probes into nanomechanical oscillators  enables ultrasensitive investigations of forces on dimensions smaller than their characteristic length scale. Meanwhile it also unravels the force field vectorial character and  how its topology impacts the measurement. Here we expose an ultrasensitive method to image 2D vectorial force fields by  optomechanically following the bidimensional Brownian motion of a singly clamped nanowire.  This novel approach relies on angular and spectral tomography of its quasi frequency-degenerated transverse mechanical polarizations: immersing the nanoresonator in a vectorial force field does not only shift its eigenfrequencies but also rotate eigenmodes orientation as a nano-compass. This universal method is employed to map a tunable electrostatic force field whose spatial gradients can even take precedence over the intrinsic nanowire properties. Enabling vectorial force fields imaging with demonstrated sensitivities of attonewton variations over the nanoprobe Brownian trajectory will have strong impact on scientific exploration at the nanoscale. \\}

\begin{figure}[t!]
\begin{center}
\includegraphics[width=0.9 \linewidth]{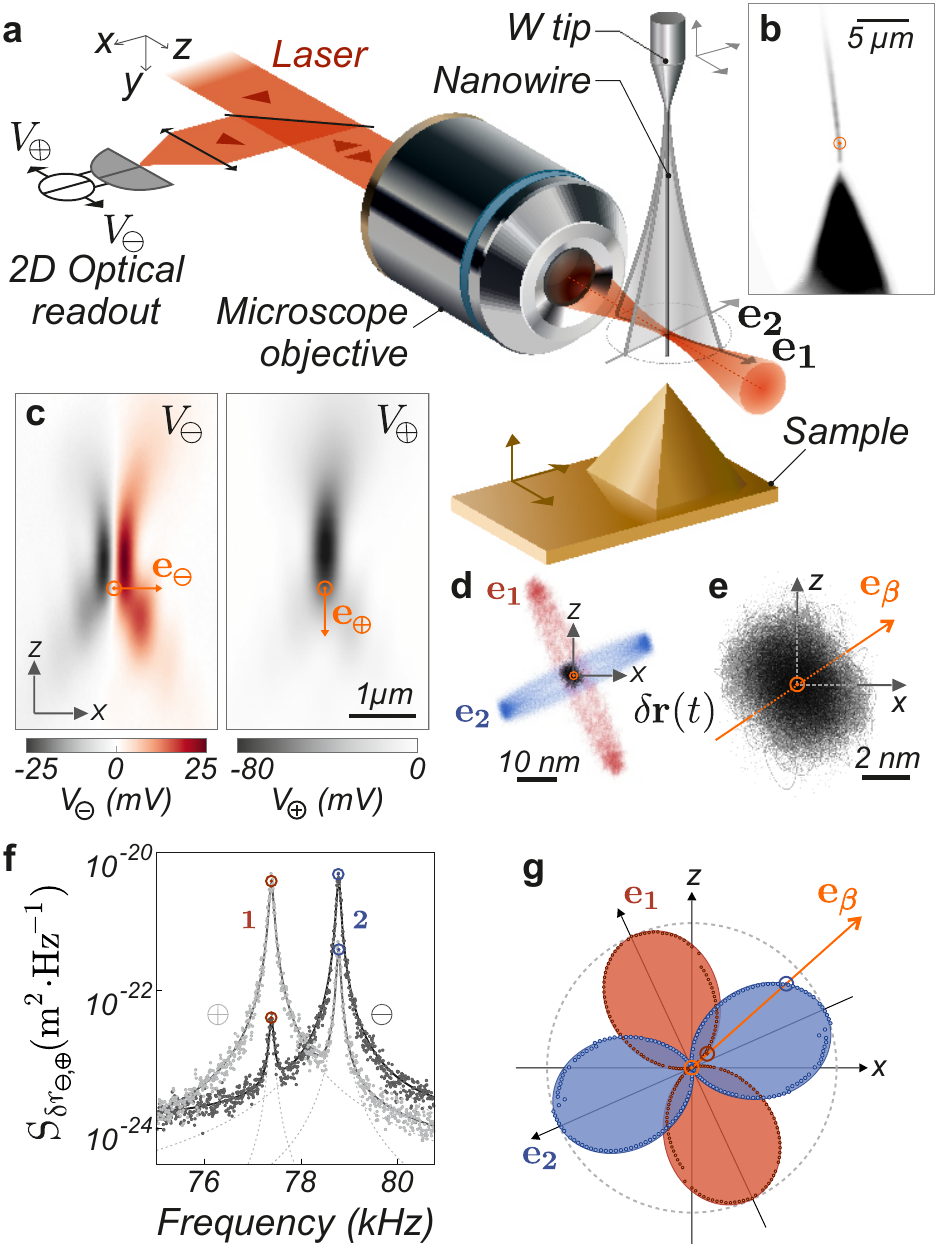}
\caption{
\textbf{ 2D optomechanical readout.} {\bf a} Schematics of the experiment. The transverse vibrations of a singly-clamped SiC nanowire are optically readout using a dual photodiode amplifier recording the reflexion of a laser beam tightly focussed on the NW extremity. {\bf b} CCD image of a voltage biased sharp metallic tip piezo-positioned at the NW extremity to generate the static electrostatic force field under investigation. {\bf c} Reflection maps of the DC differential/sum signals $V_{\oplus/\ominus} (\vv{r})$ obtained by piezo-scanning the nanowire in the laser waist. The measurement location indicated by $\odot$ provides orthogonal projective measurement vectors, $\vv{e_{\ominus,\oplus}}$ allowing a 2D reconstruction of the nanowire trajectories $\vv{\delta r (t)}$. {\bf d} Driven trajectories obtained by resonant actuation of both fundamental eigenmodes. {\bf e} Thermal noise trajectory recorded over 100\, ms (100\,ns sampling). {\bf f} Calibrated projected displacement noise spectra $S_{\delta r_{\ominus,\oplus}} [\Omega]$. {\bf g} Angular spectral tomography the 2D thermal noise, showing the relative variations of $ S_{\delta r_{\beta}} [\Omega_{1,2}]$ with the analysis angle $\beta$, giving access to the eigenmode orientations $\vv{e_{1,2}}$.}
\label{Fig1}
\end{center}
\end{figure}
{\it Introduction--}
Nanosciences were revolutionized by the invention of atomic force microscopy \cite{Binnig1986} which enabled first perceptions of the nanoworld, by measuring proximity forces exerted by a sample on a micromechanical oscillator. The subsequent emergence of nanomechanical oscillators and evolutions in  readout techniques \cite{Cleland2003,Ekinci2005,Schwab2005}  lead to impressive improvements in force sensitivity \cite{Moser2013}, enabling detection of  collective spin dynamics \cite{Degen2009,Nichol2012, Peddibhotla2013}, single electron spin \cite{Rugar2004},  mass sensing of atoms \cite{Jensen2008,Sage2015} or inertial sensing \cite{Hanay2015}. Attractive perspectives arise too when nanoresonators are hybridized to single quantum systems, such as molecular magnets \cite{Ganzhorn2013}, spin or solid states qubits \cite{Arcizet2011,Yeo2014, Montinaro2014,Pigeau2015}. This reduction of the probe size naturally motivates the exploration of force fields on dimensions smaller than their characteristic length scale where a great physical richness is expected, in particular for nanoscale imaging or investigations of fundamental interactions such as proximity forces or near field couplings. In this situation, measurements are performed in presence of strong force field gradients whose vectorial structure becomes crucially relevant and should be fully accounted to describe the measurement process itself \cite{Gloppe2014}. \\
%%%%%%%%%%%%%%%%%%%%%%%%%%%%%%%%%%%%%%%%%%%%%%%%%%%%%%%%%%%%%%%%%%%%%%%%%%%%%%%%%%%%%%%%%%%%%%%%%%%%%%%%%%%%%%%%%%%%%%%%%%%%%%%%
However, most of existing force sensing scanning probe experiments are based on oscillators such as cantilevers, whose motion remains quasi-monodimensional, even in case of multimode operation \cite{Garcia2012}. In this situation, only scalar quantities can be measured, giving access in general to the projection of the force field gradient along the measured eigenmodes which only provides partial information on the force field under investigation. \\
%%%%%%%%%%%%%%%%%%%%%%%%%%%%%%%%%%%%%%%%%%%%%%%%%%%%%%%%%%%%%%%%%%%%%%%%%%%%%%%%%%%%%%%%%%%%%%%%%%%%%%%%%%%%%%%%%%%%%%%%%%%%%%%%
In this article, we expose a novel ultrasensitive measurement technique going beyond scalar measurements and permitting to directly image vectorial 2D force fields. It is based on a singly clamped nanowire (NW) which can oscillate along two perpendicular transverse directions\cite{Siria2012,Gloppe2014,Nichol2008,Gil-Santos2010,Ramos2012,Cadeddu2016}. Its vibrating extremity is immersed in the force field under investigation while its 2D-Brownian motion is optically detected and reconstructed in realtime.
A novel measurement principle, based on the angular and spectral tomography of the NW Brownian motion gives access to the eigenfrequencies and eigenfunctions of the two quasi-degenerated fundamental vibrational modes. The latter are hybridized by the external force field gradients, which can be subsequently deduced from a direct mathematical inversion of the tomographic parameters after having understood the bidimensional coupling mechanism. The strength of the method relies on the use of quasi frequency-degenerated nanowires, whose eigenmodes are likely to be cross-coupled by the shear components of the external force field, generating large changes in eigenmode orientations as the force gradients easily take precedence over the nanowire intrinsic mechanical properties. We first introduce the principle of angular and spectral tomography (AngSTom) of the nanowire 2D thermal noise and demonstrate its capability to probe the in-plane 2D electrostatic force field gradients generated by a voltage-biased nano-tip approached in the vicinity of the nanowire extremity. The choice of  using point source-like attractive central force field permits to fully explore the panorama of dual mode coupling in 2D which is at the heart of the measurement protocol. Finally, we realize a cartography of the vectorial structure of the electrostatic force field  generated by the tip, which illustrates the strength of the method for future applications in vectorial force field scanning probe imaging.\\
%%%%%%%%%%%%%%%%%%%%%%%%%%%%%%%%%%%%%%%%%%%%%%%%%%%%%%%%%%%%%%%%%%%%%%%%%%%%%%%%%%%%%%%%%%%%%%%%%%%%%%%%%%%%%%%%%%%%%%%%%%%%%%%%
Similar NWs were previously employed \cite{Gloppe2014} to analyze the optomechanical backaction in 2D generated by a tightly focused laser field. To do so the optical force field was measured by means of a coherent pump-probe response measurement  where the laser intensity was modulated to resonantly drive the NW with optical forces. The force measurements were realized at low optical powers, so that the nanowire dynamical response was only weakly perturbed. Beyond the demonstrated novelty, this methodology was lacking of universality since the vast majority of forces under investigation in nanoscience cannot be time modulated and a novel approach which does work in presence of strong force field gradients, as expected at the nanoscale, had to be developed. The strategy consists in exploiting a quasi frequency-degenerated nanowire, whose eigenmodes get dressed by the external force field under investigation. The dimensionality of the problem requires a novel analysis of mechanical mode coupling phenomenology in 2D which is at the heart of the method exposed.

%%%%%%%%%%%%%%%%%%%%%%%%%%%%%%%%%%%%%%%%%%%%%%%%%%%%%%%%%%%%%%%%%%%%%%%%%%%%%%%%%%%%%%%%%%%%%%%%%%%%%%%%%%%%%%%%%%%%%%%%%%%%%%%%
{\it Setup--} Our nanoresonator is a $70\,\rm \mu m$-long  Silicon Carbide NW of $\simeq 200\,\rm nm$  diameter suspended at the extremity of a sharp tungsten tip. Its vibrating extremity is immersed in a strongly focused laser beam used to probe its vibrations.
The focusing is achieved by a $100\rm x/0.75$  microscope objective, producing a\, $\simeq 330\,\rm nm$ optical waist. Its long working distance ($4\,\rm mm$) bestows a convenient buffer space between the NW and the first optical element and leaves a large spatial access to the vibrating extremity of the NW for approaching the sample under test. In this work, we investigate the electrostatic in-plane force field exerted on the NW by a $17\,\rm \mu m$ high platinum-coated pyramidal tip with a $\simeq 100\,\rm nm$ radius of curvature. While the whole setup is grounded, this tip is electrically insulated from the rest of the setup  and can be voltage biased up to a few volts. The tip position can be piezo-scanned in the vicinity of the NW extremity  (see Fig.\, 1b) to realize force field cartography. The experiment is carried out in a vacuum chamber operated at a static pressure below $10^{-2}\,\rm mbar$ to suppress acoustic damping. Its temperature is stabilized within $0.01^\circ\rm C$, which limits thermal drifts below 2\,nm/h.\\

%%%%%%%%%%%%%%%%%%%%%%%%%%%%%%%%%%%%%%%%%%%%%%%%%%%%%%%%%%%%%%%%%%%%%%%%%%%%%%%%%%%%%%%%%%%%%%%%%%%%%%%%%%%%%%%%%%%%%%%%%%%%%%%%
{\it Bidimensional optomechanical readout--} An optomechanical readout is employed to fully characterize the vibration properties of the nanowire in 2D and to access the spatial structure of its Brownian motion. To do so, the light backscattered on the NW extremity is collected and focused onto an amplified split photodiode which provides the sum/difference ($V_{\oplus}$/$V_{\ominus}$) of each quadrant output voltages which permits a precise micro-positioning of the NW with respect to the laser beam in the waist area, see Fig.\,1c.
The vibrations $\vv{\delta r}(t)$ of the NW extremity in the horizontal oscillating plane around its rest position $\vv{r_0}$ are converted into temporal fluctuations in the photodetector outputs as $V_{\ominus,\oplus}(\vv{r_0}+\vv{\delta r} (t))$. Thus both signals convey a projective measurement of the NW deflection $\delta r_{\ominus,\oplus}(t)\equiv\vv{\delta r}(t)\cdot\vv{e_{\ominus,\oplus}}$ along a measurement vector defined by $\vv{e_{\ominus,\oplus}}\equiv \frac{\boldsymbol{\nabla} V_{\ominus,\oplus}}{\left|\boldsymbol{\nabla} V_{\ominus,\oplus}\right|}$ evaluated at the nanoresonator rest position $\vv{r_0}$.
By selecting a working position where both spatial gradients are at least non-collinear and ideally perpendicular (highlighted in Fig.\,1c by $\odot$), a fully 2D vectorial readout of the nanowire deflection $\delta \vv{r}(t)$ in realtime can be achieved.  Moreover, this working point lies on the optical axis which ensures the nullity of the non-conservative components of the laser dynamical backaction \cite{Gloppe2014}, which is further circumvented by working at sufficiently low optical powers ($\approx 20-60\,\rm \mu W$). Once the optical working position is determined, a 2D feedback loop is activated to stabilize the optomechanical probe against thermal drifts and static deflections (see SI). \\
%%%%%%%%%%%%%%%%%%%%%%%%%%%%%%%%%%%%%%%%%%%%%%%%%%%%%%%%%%%%%%%%%%%%%%%%%%%%%%%%%%%%%%%%%%%%%%%%%%%%%%%%%%%%%%%%%%%%%%%%%%%%%%%%
{\it Nanowire 2D thermal noise--} The 2D Brownian motion trajectory  reconstructed from the simultaneous acquisition of $\delta r_{\ominus, \oplus}(t)$ is shown in Fig.\,1d. It presents a gaussian distribution spreading over $\Delta x =\sqrt{k_B T/M_{\rm eff} \Omega_{\rm m}^2} \approx 5\,\rm nm$, which is $\approx 2.5\%$ of the NW diameter. Corresponding vibration noise spectra $S_{\delta r_{\ominus,\oplus}}[\Omega]$ are shown in Fig\,1f, restricted to the fundamental longitudinal mode family oscillating around  77\,kHz  with quality factors $Q\approx 1000$ at $5\times 10^{-5}\,\rm mbar$ and effective masses $M_{\rm eff} = 600\,\rm fg$ (stiffness  $k_i\approx 150\,\rm \mu N/m$). A slight asymmetry in the NW geometry lifts the frequency degeneracy between the two mechanical polarizations to $\Omega_2-\Omega_1\approx 2\pi\times 400\,\rm Hz$ whose orientations  $\vv{e_{1,2}}$ can be visualized in the 2D representation under resonant actuation of the NW, see Fig.\,1d. To fully exploit the NW sensing capacity, it is necessary to analyze its 2D thermal noise and detect modifications of effective temperatures, eigenfrequencies and eigenmode orientations. To do so the recorded 2D NW thermal noise trajectory  $\vv{\delta r} (t)$ is numerically projected along an arbitrary unitary measurement vector $\vv{e_{\beta}}$, providing the projected displacement  $\delta r_\beta(t)\equiv \vv{\delta r}(t)\cdot \vv{e_{\beta}}$, whose spectral density  is computed and evaluated at the eigenfrequencies yielding $S_{\delta r_{\beta}}[\Omega_{1,2}]$. Repeating this sequence for varying measurement angles allows to reconstruct the angular and spectral tomography of the nanowire thermal noise in 2D, as shown in Fig.\,1g. This representation permits an identification of the eigenmode orientations $\vv{e_{1,2}}$ as measurement directions which maximize the noise peak heights and an estimation of the NW effective masses and temperatures, see SI.\\

\begin{figure}[t!]
\begin{center}
\includegraphics[width=0.99\linewidth]{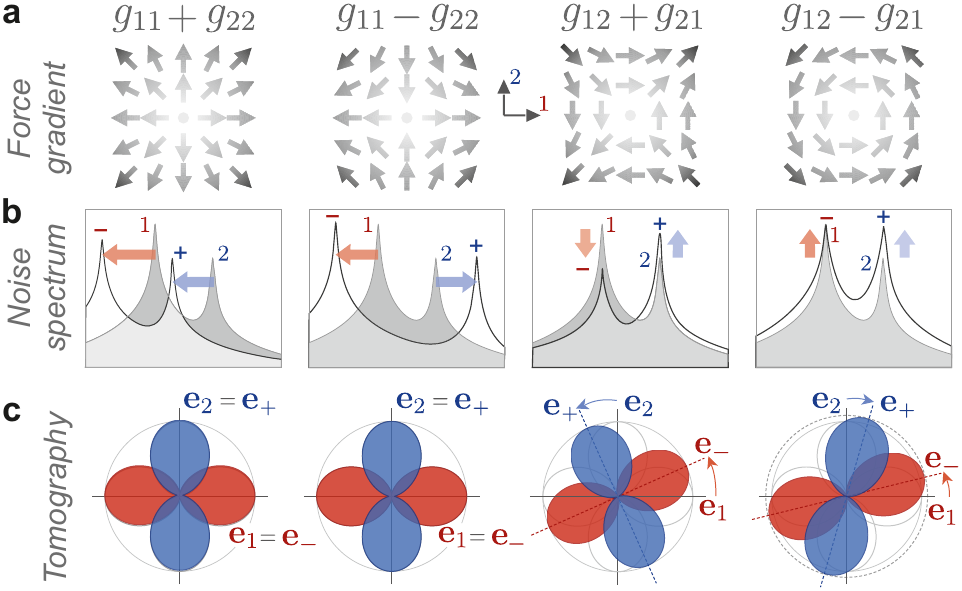}
\caption{
\textbf{ Measurement principle of 2D force field gradients.} Any 2D force field gradients can be fully expanded on the geometric basis represented in {\bf a}. Their respective impact on  thermal noise spectra (measured along one single measurement vector) and on ASTom representations are shown in {\bf b}  and {\bf c}. The knowledge of the modified eigenvectors' orientations and eigenfrequencies allows a complete derivation of the local external force field gradients.}
\label{Fig2}
\end{center}
\end{figure}

\newsavebox{\smlmat}% Box to store smallmatrix content
\savebox{\smlmat}{$\left(\begin{smallmatrix}\Omega_1^2&0\\
0& \Omega_2^2\\
\end{smallmatrix}\right)$}

{\it 2D force field imaging--} The modification of the nanowire 2D thermal noise in presence of an external force field gradient is at the heart of the measurement principle, which is now exposed in detail. The NW dynamics, restricted to the 2 fundamental eigenmodes,  is described by a 2D point-like oscillator dynamics:
\begin{equation}
\vv{ \delta \ddot r}=
- \boldsymbol{\Omega^2}\cdot \vv{\delta r}
-\Gamma \vv{ \delta \dot r}
+\frac{ \vv{ F}({\bf r_0}+\vv{ \delta r})}{M_{\rm eff}}
+\frac{\vv{\delta F_{\rm th}}}{M_{\rm eff}},
\end{equation}
where $\boldsymbol{\Omega^2}\equiv$~\usebox{\smlmat} is the intrinsic restoring force matrix in the $\vv{e_1},\vv{e_2}$ basis, $\Gamma$ the mechanical damping rate, $M_{\rm eff}$  their effective mass and  $\vv{\delta F_{\rm th}}$ represents the 2D Langevin force vector, independently driving each eigenmode  \cite{Pinard1999} with a white force noise of spectral density $S_F^{\rm th}= 2 M_{\rm eff}\Gamma k_B T$. The use of the normal mode expansion is justified by the low ($\lesssim 10\%$) damping asymmetry observed on the uncoupled eigenmodes \cite{Schwarz2016}.
$\vv{F}(\vv{r_0}+\vv{\delta r})$ represents the action of the force field under investigation on the two fundamental eigenmodes, evaluated at the vibrating extremity position. In this work we investigate situations where the force field experienced by the NW (here of electrostatic origin) instantaneously follows -on mechanical timescales-  the change in the NW position. Delays in the establishment of the force \cite{Barois2012,Ramos2012,Nigues2015} could be treated with the same formalism  \cite{Gloppe2016}.\\
The force experienced by the NW can be linearized as $ \vv{ F}({\bf r_0}+\vv{ \delta r})\approx \vv{F}(\vv{r_0})+\left.\left(\vv{\delta r}\cdot\boldsymbol{\nabla}\right)\vv{F}\right|_{\vv{r_0}}$. The first term represents the mean force undergone by the NW, responsible for a static deflection which is compensated by the 2D position feedback maintaining the NW extremity at the measurement location \cite{Rodrigues2012}. As a side note, recording the correction of the feedback loop in both directions $\vv{\delta r}_{\rm stat}$ for varying positions of the sample under test (see Fig.\,4c) enables a first static method to probe the 2D in-plane vectorial force field according to $ \vv{F}= {M_{\rm eff}} \boldsymbol{\Omega^{2}}\cdot \vv{\delta r}_{\rm stat}$. A static force larger than $\sqrt{ M_{\rm eff} \Omega_{\rm m}^2 k_B T}\approx  1\,\rm pN$ will induce a nanowire deflection exceeding the spatial spreading of its thermal noise. As described below, a dynamical measurement of the NW Brownian motion around its rest position allows for a much more sensitive and robust sensing.\\
\begin{figure}[t!]
\begin{center}
\includegraphics[width=0.99\linewidth]{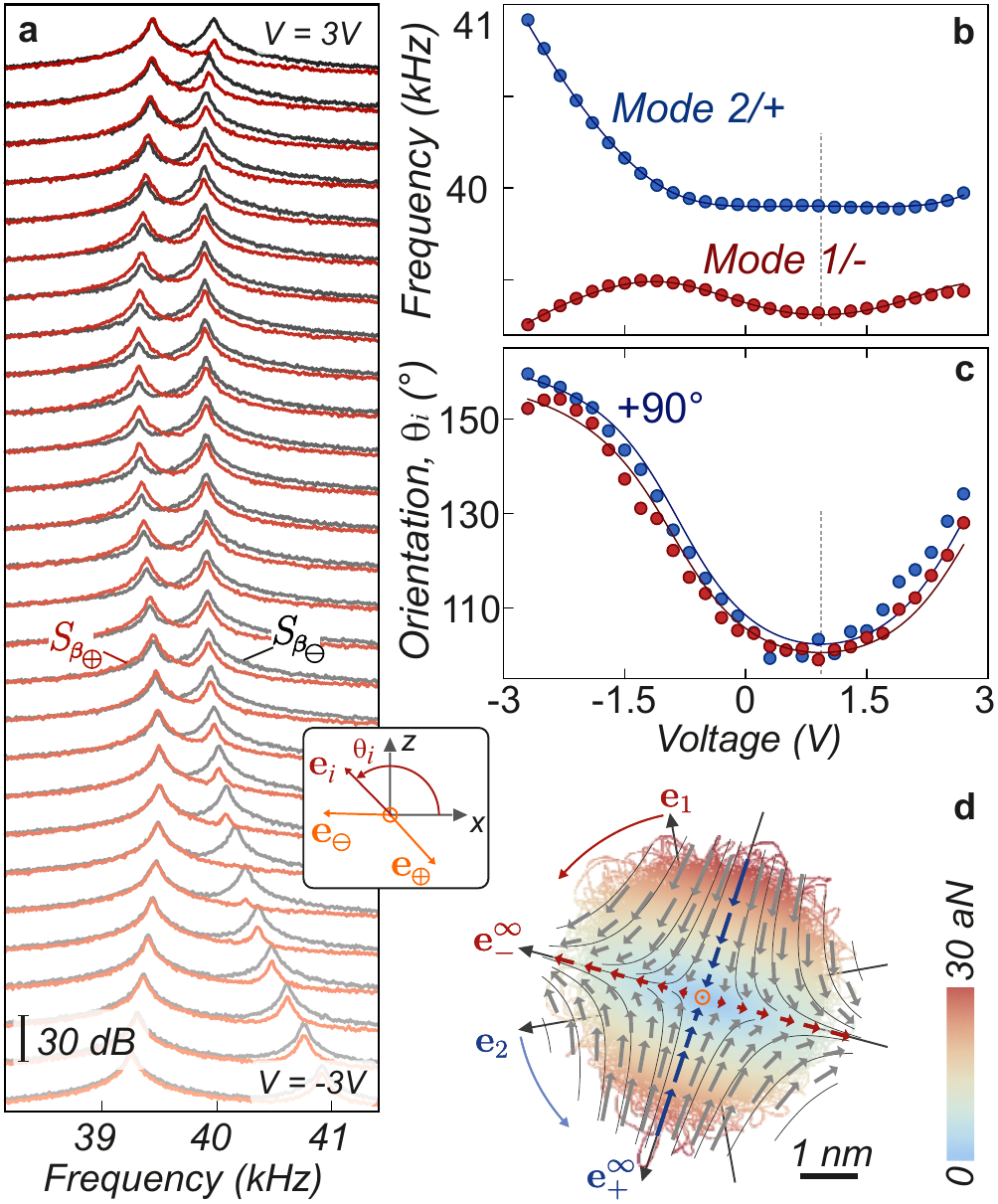}
\caption{
\textbf{ Mapping electrostatic force field gradients}  A voltage bias is applied on the electrostatic tip to tune the external force field gradient experienced by a $\approx 40\,\rm kHz$ NW for a nanotip position chosen to generate large cross-coupling between eigenmodes. {\bf a} Thermal noise spectra  $S_{\delta r_{\ominus,\oplus}}[\Omega]$ obtained for varying voltage bias. {\bf b,c} Deduced frequency shifts and eigenmode rotations whose knowledge permits a full derivation of the force field gradients $\partial_iF_j$. Solid lines are fits according to the theory, based on a quadratic dependence of the electrostatic force field, using $ M_{\rm eff} g_{ij}= \{-146, 100, 114, 14\}\,\rm nN/m/V^2$ (i,j=1,2)  and an offset voltage of 1 V, see text and SI.  {\bf d} Linearly extrapolated cartography of the local force field ($\vv{F}(\vv{r})-\vv{F}(\vv{r_0})$) drawn over the spatial extension of the NW thermal noise for a 300\, mV bias. At large bias, the NW eigenvectors get aligned with the eigenvectors $\vv{e_\infty^\pm}$ of the electrostatic force field gradient matrix $g_{ij}$.}
\label{Fig3}
\end{center}
\end{figure}
The second linear term in the force field expansion is responsible for a modification of the restoring forces experienced by the nanowire, which affects its dynamics and the spatial architecture of its Brownian motion. In Fourier space, the NW deflection can be expressed as $\vv{\delta r}[\Omega]= \boldsymbol{\chi}[\Omega]\cdot \vv{\delta F_{\rm th}}[\Omega]$
where we have introduced the mechanical susceptibility matrix, whose inverse reads in the unperturbed basis $\vv{e_{1,2}}$:
\begin{equation}
\boldsymbol{\chi^{-1}}[\Omega]=\left(\begin{array}{cc}
\chi_1^{-1}[\Omega]-g_{11}&-g_{21}\\
-g_{12}& \chi_2^{-1}[\Omega]-g_{22}\\
\end{array}\right)
\end{equation}
where $\chi_{1,2}^{-1}\equiv M(\Omega_{1,2}^2-\Omega^2-i\Omega \Gamma)$  are the unperturbed mechanical susceptibilities. Due to the immersion in the external force field, the NW dynamics now depends on the 4 gradients of the 2D force field:
$ g_{ij} (\vv{r_0})\equiv\frac{1}{M_{\rm eff}} \left.\partial_i F_j\right|_{\vv{r_0}} $
whose shear components ($i\neq j$) control the cross-coupling between eigenmodes.
The susceptibility matrix is diagonalized into new eigenmodes of vibration, labelled with $\pm$ indices with eigenfrequencies:
\begin{equation}
\Omega_{\pm}^2\equiv\frac{\Omega_{1\parallel}^2+  \Omega_{2\parallel}^2}{2}\pm
\frac{1}{2}\sqrt{
\left(\Omega_{2\parallel}^2-\Omega_{1\parallel}^2\right)^2+ 4g_{12} g_{21}
}
\end{equation}
and unitary eigenvectors:
\newsavebox{\toto}% Box to store smallmatrix content
\savebox{\toto}{$\left(\begin{smallmatrix}\Delta \Omega_\perp^2\\ g_{12}\\
\end{smallmatrix}\right)$}
\newsavebox{\tata}% Box to store smallmatrix content
\savebox{\tata}{$\left(\begin{smallmatrix}-g_{12}\\ \Delta \Omega_\perp^2\\
\end{smallmatrix}\right)$}
${\vv{ e_-}}\equiv \frac{1}{\sqrt{g_{12}^2+\Delta \Omega_\perp^2}}$~\usebox{\toto} and
${\vv{ e_+}}\equiv \frac{1}{\sqrt{g_{21}^2+\Delta \Omega_\perp^2}}$~\usebox{\tata}
where we have introduced $\Omega_{i\parallel}^2\equiv\Omega_{i}^2-g_{ii}$ and $\Delta \Omega_\perp^2\equiv \Omega_{2\parallel}^2-\Omega_-^2=\Omega_+^2-\Omega_{1\parallel}^2$.
The impact of the force field gradients on the NW thermal noise is illustrated in Fig.\,2 in a simplified picture, while the full calculation of the 2D thermal noise spectra is carried out in the SI. The force field divergence $g_{11}+g_{22}$ is responsible for a global frequency shift, while Poisson contributions $g_{11}-g_{22}$ modify the frequency splitting. On the contrary, shear components  $g_{i\neq j}$ are responsible for rotation of eigenmode orientations. In a conservative force field, verifying $g_{12}=g_{21}$, both eigenmodes are equally rotated, which preserves their initial orthogonality  since $\vv{e_{-}}\cdot\vv{e_+}\propto \vv{rot}(\vv{F})\cdot \vv{e_y}$.  The situation is qualitatively different in a non-conservative force field which breaks eigenmode orthogonality and increases their effective temperatures, see SI for full calculation.
This last observation implies that a 2D vectorial readout is essential to fully determine a 2D force field. Indeed a measurement based on a single projective readout is unable to distinguish between eigenmode rotation and effective temperature increase.
As can be mathematically derived, see SI, it is sufficient to determine the eigenfrequencies ($\Omega_\pm$) and eigenvectors($\vv{e_\pm}$) of both fundamental eigenmodes to fully determine the 4 components of the force field gradients. Those are precisely the quantities that can be extracted from the spectral and angular tomography method introduced above, see SI.
\begin{figure*}[t!]
\begin{center}
\includegraphics[width=0.95\linewidth]{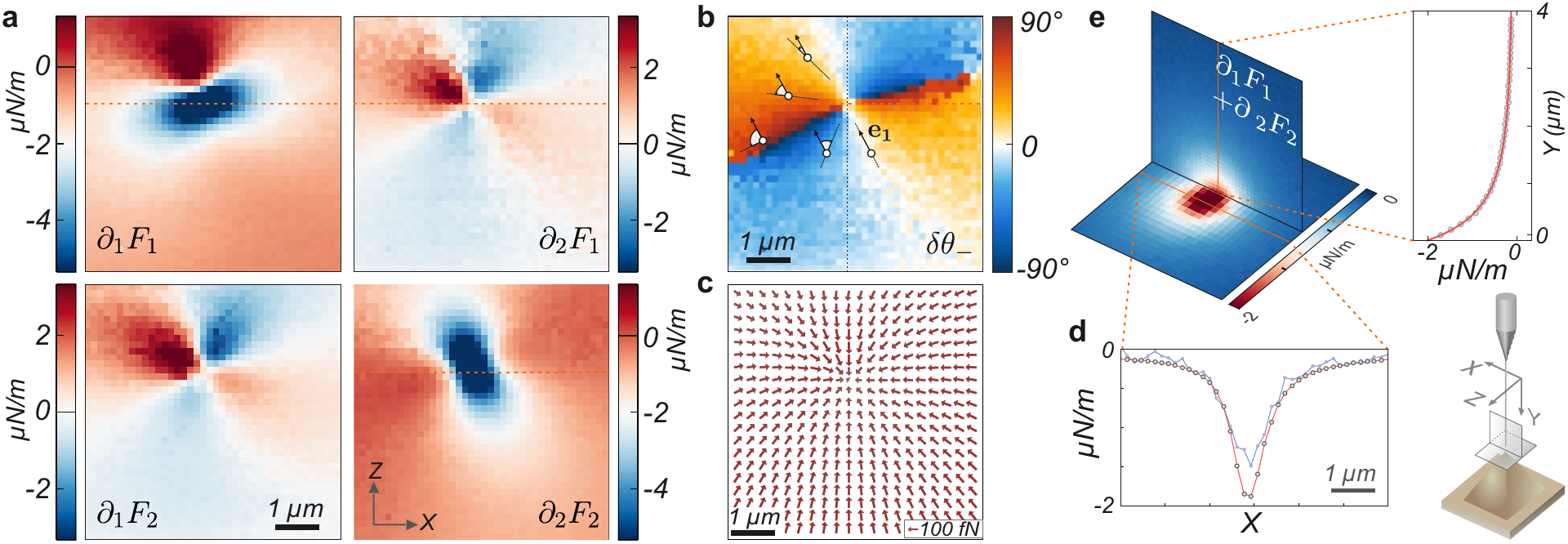}
\caption{
\textbf{ Scanning probe measurements of electrostatic force field gradients.} {\bf a} Map of the force field gradients obtained for a bias voltage of -2\,V while scanning the tip $\approx 500\,\rm nm$ below the nanowire extremity in a $6\times 6\,\rm \mu m^2$ area (150\,nm grid). {\bf b}  Map of relative eigenmode rotation ($\delta\theta_-$) illustrating the nano-compass mechanism: the $\vv{e_1/-}$ eigenmode is pointing towards the softer electrostatic gradient, which is generally oriented towards the nanotip. ($1\,\rm \mu m$ scalebar). {\bf c} Map of the 2D force field computed by integrating the measured 2D gradients for 1\,V bias, using an integration constant (force vector of (-37, +4) fN) which cancels the force field just above the tip, see SI. The maximum static deflection measured was of $\approx 4\,\rm nm$, see SI, which is small compared to the characteristic length over which the force field varies. {\bf d} Comparison of static  (blue) and dynamical  (red) measurements of the force field divergence ($\partial_xF_x+\partial_zF_z$). {\bf e} Vertical cartography of the force field gradients (divergence and shear components) measured for a 1\,V bias. Inset: the minimum in-plane divergence of the force field is well adjusted (red line) with a $y^{-3}$ power law, see SI.}
\label{Fig4}
\end{center}
\end{figure*}

{\it Probing electrostatic force field gradients--}
To illustrate the strength of the method and demonstrate its validity with a well controlled force field, we employ the nanowire to probe the electrostatic force field generated by the metallic tip. The latter is piezo-positioned at a location where large shear components are generated to enhance eigenmodes hybridization. Thermal noise spectra obtained for varying bias voltage $V$ applied on the tip are shown in Fig.\,3a while the eigenfrequencies and eigenvectors orientations ($\vv{e_\pm}\cdot \vv{e_x}$) are reported in Fig.\,3b and 3c respectively.
Both eigenmodes rotate while preserving their initial orthogonality which is a direct experimental verification of the conservative character of the electrostatic force field.
Strikingly, a rotation of almost $60^\circ$ can be achieved at large biases, indicating that the external force field prevails over the intrinsic NW properties to determine its eigenmode structure.
The force field gradients are subsequently extracted from eigenfrequencies and eigenmode orientations and this for each bias voltage, permitting to verify their quadratic scaling according to  $g_{ij}= g_{ij}^0+ \alpha_{ij} (V-V_0)^2$, as expected for electrostatic actuation, see Fig.\,3.  The very good agreement observed all across the bias range demonstrates the validity of the method even in extremely large force field gradients.
The corresponding interpolated force field ($M_{\rm eff} \, g_{ij} \delta r_i $) evaluated for a bias voltage of $V-V_0=300\,\rm mV$ is shown in Fig.\,3d. The extreme sensitivity of the NW is revealed: the relative variations of the force field magnitude over the spatial spreading of its thermal noise are only of  30\,aN. The intrinsic low/high frequency eigenmodes (1/2)  are dressed by the external force field, defining the new  set of eigenmodes (-/+) which tend to align their orientation with the eigenaxes of the external force field gradient matrix. The eigenvectors $\vv{e_\infty^\pm}$ of the $g_{ij}$ matrix are also shown and they indeed match the eigenmodes' orientations measured at large electrostatic biases.
This representation also permits understanding the change of curvature visible in the frequency shifts (see SI). For example at low bias voltages ($\lesssim 300\,\rm mV$) the low frequency mode (1/-) still oscillates along its intrinsic orientation $\vv{e_1}$ and experiences a trapping force field pointing towards the center, which results in a frequency increase. However at larger bias, it rotates and finally gets aligned with $\vv{e_\infty^-}$, where it experiences an anti-trapping force field, which lowers its oscillation frequency.
This dynamical electrostatic force measurement \cite{Martin1988,Terris1989} also permits to access the local electrostatic zero, $V_0\approx 1\,\rm V$ here, whose spatial variations play an essential role in surface quality analysis and proximity force measurements \cite{Behunin2012,Siria2012}.\\
Qualitatively, the longitudinal force field gradient required to shift eigenfrequencies by their mechanical linewidths amounts to $\partial_1 F_1= k_1/Q \approx 10^{-7}\,\rm N/m$, while the shear force field gradients required to rotate eigenmodes by $45^\circ$ amounts to $\partial_2 F_1= k_1 (\Omega_2-\Omega_1)/\Omega_1 \approx 2\times 10^{-7}\,\rm N/m$ for a $0.5 \% $ intrinsic frequency splitting and a quality factor of 1000. These expressions underline the importance to work with soft NWs, presenting large aspect ratios, large quality factors and weak intrinsic asymmetries to achieve large sensitivities to both longitudinal and transverse force field gradients.

{\it Cartography of electrostatic force field gradients--}
The force field gradients  generated by the nanotip can be spatially mapped by piezo-scanning its position with respect to the nanowire extremity, see Fig.\,4.  The electrostatic tip was horizontally scanned over $40\times 40$ positions in a $6\times 6\,\rm\mu m^2$ area at a vertical distance of $\approx 500\,\rm nm$ below the NW extremity. In each position, the bias voltage was scanned between -2\,V and 2\,V while simultaneously recording the 2D thermal noise and analyzing modifications in frequency, orientations and static deflection (3\,s acquisition time).  The map of the low frequency mode rotation angle is shown in Fig.\,4b, illustrating the nano-compass effect as this mode always rotate towards the softer direction( $\vv{e_\infty^-}$), keeping track of the nanotip position.
Reproducing in each point the analysis exposed above, the spatial map of the force field gradients $\partial_{i} F_j (\vv{r}),\, i,j=1,2$ can be reconstructed as shown in Fig.\,4a. For our quasi point-like attractive central force field under investigation, unperturbed eigenmode orientations appear as symmetry axes in the divergence-like components ($\partial_iF_i$) maps and as anti-symmetry axes for shear components.  Shear component maps are identical (within experimental uncertainties) as expected from the conservative nature of the electrostatic actuation ($\partial_1 F_2=\partial_2 F_1$). A tilt in the tip scanning plane with respect to the NW axis is responsible for the slight top/bottom asymmetry observed.
A phenomenological evaluation based on a radial approximation \cite{Derjaguin1958} is in good agreement with our measurements (see SI).
The dynamically measured 2D force field gradients can be compared to the spatial gradients measured from of the static NW deflexion map (see SI): Fig\,4d compares the force field 2D divergence  $\propto g_{11}+ g_{22}$ obtained from both methods. While both results are in quantitative agreement,  as expected dynamical measurements give by far a larger measurement quality.
The NW extremity can also be scanned in the vertical plane and thanks to its large sensitivity, force field gradients can still be measured microns away from the surface, see Fig. 4e.
Then the 2D force field can be reconstructed through a 2D integration of the 4 force field gradients maps, including a constant force vector chosen to cancel the force field above the tip, see SI. Its attractive pattern is shown in Fig.\,4c, allowing for a detailed investigation of the electrostatic behaviour of EFM tips.
Beyond demonstrating the scanning probe imaging capacity of the system,  this measurement also permits exploring the great richness of strong dual mode coupling in 2D: the mode crossing/anti-crossing phenomenology \cite{Faust2013} can be tuned by carefully positioning the tip position, see SI. This also permits engineering the 2D force probe to maximize its sensitivity to specific components of the force field gradients. As an example, its sensitivity to shear component can be maximized  by electrostatically minimizing its intrinsic frequency splitting.
The measurements accumulated at varying distances (both horizontally and vertically) and bias voltages have permitted verifying the linearity of our probe on a large dynamical range spanning from a few nN/m to tens of $\rm \mu N/m$.\\

{\it Conclusions--}
We have demonstrated a novel approach allowing vectorial force field imaging at the nanoscale, especially suited for probing in-plane 2D force fields. By bringing an additional dimension to surface imaging techniques based on mechanical signals with demonstrated force sensitivities in the attoNewton range at room temperature -several orders of magnitude better than commercial AFM apparatus- we anticipate that our approach will quickly bring novel perspectives in scientific imaging at the nanoscale.  It will contribute to shed new light on the field of probe-surface interaction \cite{Mate1987,Karrai2000,Kuehn2006,Siria2012}. Developments of novel measurement protocols based on driven 2D trajectories will permit both increasing the measurement speed and improving the sensitivity, while simultaneous observation of several longitudinal modes families \cite{Garcia2012} will further enrich the imaging capacity of our vectorial force probe.
In particular, it could be straightforwardly employed to explore proximity forces at the nanoscale, such as Casimir forces in nano-structured samples where novel phenomenology can be expected \cite{Rodriguez2011,Chen2002,Guerout2015}. The method is also compatible with non-conservative force field imaging and will permit further exploration of fluctuation theorems in 2D. Finally, our 2D analysis can also be transposed to 3D, using 3D angular and spectral analysis with for instance trapped ions \cite{Bushev2013} or levitated nano-particles \cite{Raizen2011,Gieseler2012}.
%%%%%%%%%%%%%%%%%%%%%%%%%%%%%%%%%%%%%%%%%%%%%%%%%%%%%%%%%%%%%%%%%%%%%%%%%%%
%%%%%%%%%%%%%%%%%%%%%%%%%%%%%%%%%%%%%%%%%%%%%%%%%%%%%%%%%%%%%%%%%%%%%%%%%%%
%%%%%%%%%%%%%%%%%%%%%%%%%%%%%%%%%%%%%%%%%%%%%%%%%%%%%%%%%%%%%%%%%%%%%%%%%%%
%%%%%%%%%%%%%%%%%%%%%%%%%%%%%%%%%%%%%%%%%%%%%%%%%%%%%%%%%%%%%%%%%%%%%%%%%%%

\textit{Acknowledgements---} We warmly thank the PNEC group at ILM, A. Gloppe, B.\,Canals, J.\, Chevrier, S.\, Reynaud, A.\, Lambrecht, J.P.\, Poizat, G.\, Bachelier, J.\,Jarreau, C.\,Hoarau, E.\,Eyraud and D.\,Lepoittevin, for theoretical, experimental and technical assistance. This project is supported by the  ANR (FOCUS-13-BS10-0012), the ERC Starting Grant StG-2012-HQ-NOM and Lanef (CryOptics).

\bibliographystyle{naturemag}
%\bibliography{Article-Force}

\end{document}